\documentclass[sn-mathphys,Numbered]{sn-jnl}

\usepackage{graphicx}%
\usepackage{multirow}%
\usepackage{amsmath,amssymb,amsfonts}%
\usepackage{amsthm}%
\usepackage{mathrsfs}%
\usepackage[title]{appendix}%
\usepackage{xcolor}%
\usepackage{textcomp}%
\usepackage{manyfoot}%
\usepackage{booktabs}%
\usepackage{algorithm}%
\usepackage{algorithmicx}%
\usepackage{algpseudocode}%
\usepackage{listings}%
\usepackage{amsfonts,amssymb,stmaryrd,latexsym,amsmath,braket}
\usepackage{graphicx,subfigure}
\usepackage{times}
\usepackage{slashed}
\usepackage{braket}
\usepackage{verbatim}
\usepackage[utf8]{inputenc}
\usepackage{soul}

\usepackage{color}
\usepackage{lipsum,adjustbox}
\usepackage[capitalise]{cleveref}
\usepackage{tikz}
\usetikzlibrary{quantikz}

\theoremstyle{thmstyleone}%
\theoremstyle{thmstyletwo}%

\theoremstyle{thmstylethree}%

\begin{document}

\title[Article Title]{Quantum techniques for eigenvalue problems}


\author[*]{\fnm{Dean} \sur{Lee}}\email{leed@frib.msu.edu}

\affil[*]{\orgdiv{Facility for Rare Isotope Beams and Department of Physics and Astronomy}, \orgname{Michigan State University}, \orgaddress{\city{East Lansing}, \state{MI} \postcode{48824}, \country{USA}}}


\abstract{This article is a brief introduction to quantum algorithms for the eigenvalue problem in quantum many-body systems.  Rather than a broad survey of topics, we focus on providing a conceptual understanding of several quantum algorithms that cover the essentials of adiabatic evolution, variational methods, phase detection algorithms, and several other approaches.  For each method, we discuss the potential advantages and remaining challenges.}




\maketitle

\section{Introduction}\label{sec1}

Quantum computing has the potential to address many of the unsolved problems of quantum many-body physics. By allowing for arbitrary linear combinations of tensor products of qubits, one can store
exponentially more information than classical bits.  This opens the possibility of calculations of strongly-interacting systems with many degrees of freedom without the
need for Monte Carlo methods and their accompanying problems associated with
sign oscillations \cite{loh1990sign,henelius2000sign,Alhassid:2014fca,Lahde:2015ona,Frame:2017fah,Lonardoni:2018nob}.  Furthermore, qubits
naturally evolve with unitary real-time dynamics, providing access to non-equilibrium processes, which are often well beyond the reach of first-principles
calculations using classical computers. But there are also great challenges to realizing the promise of quantum computing. One of the main problems is the fact that the quantum computing devices available today have significant limitations due to gate errors, qubit decoherence, faulty measurement readout, small numbers of qubits, and limited qubit connectivity. These problems severely limit the class of problems that one can address at present.  Nevertheless, significant advances are being made in quantum hardware performance and scale \cite{chow2021ibm, pelofske2022quantum, neyenhuis2023quantinuum}, and it is useful to consider the design and performance of quantum algorithms as quantum resources grow and become more reliable.  

There is an excellent and comprehensive review on quantum computing and quantum many-body systems in Ref.~\cite{Ayral:2023ron}.  Instead of writing another review with similarly broad scope, in this article we instead focus on several algorithms of interest for eigenvalue problems.  The aim is to provide a readable introduction for novice readers with enough detail to demonstrate the concepts and execution of each method.  We should note that there are many useful algorithms of relevance to eigenvalue problems that we do not cover here.  These include cooling algorithms \cite{boykin2002algorithmic,rempp2007cyclic,Lee:2019zze,Gustafson:2020vqg,Ball:2022dxy}, coupled heat bath approaches \cite{Boykin:2002,Xu:2014}, dissipative open system methods \cite{Kraus:2008,Verstraete:2009,sinayskiy2012efficiency}, spectral combing
\cite{Kaplan:2017ccd}, symmetry projection techniques \cite{yen2019exact,Siwach:2021krs,RuizGuzman:2021qyj,seki2022spatial,Lacroix:2022vmg}, linear combinations of unitaries \cite{2012arXiv1202.5822C,Roggero:2020sgd,Lv:2022hjz,Siwach:2022ugy,meister2022tailoring}, and imaginary time evolution \cite{Motta:2019,mcardle2019variational,yeter2020practical,Turro:2021vbk,Jouzdani:2022yrf}.

In the following, we start with a review of the adiabatic theorem and the performance of adiabatic evolution for the preparation of eigenstates.  After this, we cover the broad class of variational methods. We discuss gradient calculation techniques for optimization and several specific variational algorithms.  Thereafter we present several phase detection algorithms.  These include phase estimation, iterative phase estimation, and the rodeo algorithm. We then conclude with a summary and outlook for the future.  

\section{Adiabatic Evolution}

The adiabatic theorem states that if a quantum state is an eigenstate of an initial Hamiltonian $H(0) = H_0$, then the quantum state will remain trapped in an exact eigenstate of the instantaneous Hamiltonian $H(t)$ in the limit that the time dependence of $H(t)$ is infinitely slow \cite{Born:1928,Kato:1950}.  If this evolution has only finite duration, then the error will scale inversely with the total time evolution, $T$.
We can use quantum adiabatic evolution to prepare the eigenstates of any Hamiltonian $H_1$ by preparing an exact eigenstate of some simple initial Hamiltonian $H_0$.  For the purpose of analysis, it is convenient to scale out the dependence on the total duration of time $T$ and work with the rescaled variable $s = t/T$.  We then make a smooth interpolation $H(s)$ with $s$ ranging from $s= 0$ to $s=1$, with $H(0)=H_0$ and $H(1)=H_1$  \cite{Farhi:2000,doi:10.1126/science.1057726,childs2001robustness,van2001powerful,farhi2002quantum,roland2002quantum,santoro2006optimization,Wiebe:2012,richerme2013experimental,RevModPhys.90.015002,CoelloPerez:2021jkh,vanVreumingen:2023aei}. Let us define the adiabatic evolution operator
\begin{equation}
    U(s) = {\cal T} \exp \left[-i T\int_0^{s} H(s') ds' \right], \label{evol} 
\end{equation}
where ${\cal T}$ indicates time ordering where operators at later times are placed on the left.  In the limit of large time $T$, the unitary transformation $U(1)$ will map any eigenstate of $H_0$ to an eigenstate of $H_1$.  In Ref.~\cite{Elhatisari:2022qfr}, it is observed that the unitarily-transformed Hamiltonian,
\begin{equation}
    H'(1) = U^\dagger(1)H_1 U(1),
\end{equation}
is a Hamiltonian whose eigenvalues are equal to $H_1$ but whose eigenvectors are equal to $H_0$.  For this reason, the term ``Hamiltonian translator'' was used to describe the unitary transformation $U(1)$.  Suppose we start from the Hamiltonian $H(0)$ and perform a perturbation theory expansion in the difference, $H'(1)-H(0)$,
\begin{equation}
  H'(1) = H(0) + [H'(1)-H(0)].
\end{equation}
Since $H(0)$ and $H'(1)$ share the same eigenvectors, we find that first-order perturbation for the energy is exact and all other terms in perturbation theory for the energy or wave function vanish.  

Let us now consider the one-parameter eigenvector $\ket{\psi(s)}$, which is an instantaneous eigenvector of $H(s)$ for $s$ in the interval $[0,1]$.  Let $\Delta(s)$ be the spectral gap between $\ket{\psi(s)}$ and the rest of the energy spectrum of $H(s)$.  In computing the spectral gap, we can ignore sectors that are orthogonal to $\ket{\psi(s)}$ due to symmetries that are respected by $H(s)$.  We note that $\ket{\psi(0)}$ is an eigenstate of $H_0$.  Let us define $\ket{\psi_U(s)}$ as $U(s)\ket{\psi(0)}$.  

We use the symbol $|| \cdot ||$ to denote the operator norm.  Building upon the work of Ref.~\cite{avron1987adiabatic,avron1993adiabatic,klein1990power}, Jansen et al. \cite{Jansen:2007} derived the rigorous bound that
\begin{equation}
   T \geq \frac{1}{\delta}\left\{ \int^s_0 \left[ \frac{|| \partial_s^2 H(s')||}{\Delta^2(s)} + 7 \frac{|| \partial_s H(s')||^2}{\Delta^3(s)}\right] ds' + B \right\} \label{AE}
\end{equation}
is sufficient to satisfy the error bound
\begin{equation}
    |\braket{\psi(s)|\psi_U(s)}| \geq 1-\delta,
\end{equation}
where $B$ is a boundary term that vanishes when $\partial_s H(0)$ and $\partial_s H(1)$ both equal zero \cite{vanVreumingen:2023aei}.  We see from Eq.~\eqref{AE} that, for any fixed system, the required time $T$ is scaling inversely with the error $\delta$. 

The challenge with adiabatic state preparation for eigenstates of quantum many-body systems is the fact that $\Delta(s)$ may be extremely small for large systems.  This is especially true when $H_0$ and $H_1$ have very different eigenstates, and $H(s)$ must pass through one or more quantum phase transitions.  This motivates the search for initial Hamiltonians $H_0$ for which the starting eigenstate can be prepared on a quantum computer, but the eigenstate structure of $H_0$ is not completely trivial and has some resemblance to that of $H_1$ \cite{Sarkar:2023qjn}.  Even in cases where $\ket{\psi_U(1)}$ is not a good approximation to the eigenstate $\ket{\psi(1)}$ of $H_1$, the state $\ket{\psi_U(1)}$ can still be a useful starting vector for other state preparation algorithms which converge more rapidly.

In order to perform the time evolution in Eq.~\eqref{evol}, one usually uses some version of the Trotter approximation.  The conceptual starting point for the Trotter approximation is the Baker-Campbell-Hausdorff formula, which states that when $e^Ae^B = e^C$, we have the formal series
\begin{equation}
    C = A + B + \frac{1}{2}[A,B] + \frac{1}{12}[A,[A,B]] - \frac{1}{12}[B,[A,B]] + \cdots.
\end{equation}
If our Hamiltonian has two non-commuting pieces,
\begin{equation}
    H = H_A + H_B,
\end{equation}
then, at first order in the Trotter-Suzuki expansion, we can use \cite{trotter1959product}
\begin{align}
e^{-iH\Delta t} &= e^{-iH_A\Delta t}e^{-iH_B\Delta t} + O[(\Delta t)^2] \nonumber \\
&= e^{-iH_B\Delta t}e^{-iH_A\Delta t} + O[(\Delta t)^2].
\end{align}
At second order we have
\begin{align}
    e^{-iH\Delta t} & = e^{-iH_B\Delta t/2}e^{-iH_A\Delta t}e^{-iH_B\Delta t/2} + O[(\Delta t)^3] \nonumber \\
    & = e^{-iH_A\Delta t/2}e^{-iH_B\Delta t} e^{-iH_A\Delta t/2}+ O[(\Delta t)^3].
\end{align}
The generalization to higher-order expressions can be found in Ref.~\cite{Suzuki:1993}.  The performance of the Trotter-Suzuki expansion can be improved in numerous ways, such as using random orderings \cite{childs2019faster,campbell2019random}, sums of Trotter products at different orders \cite{low2019well,watkins2022time,faehrmann2022randomizing}, extrapolation methods \cite{rendon2022improved}, and renormalization \cite{carena2021lattice}.

\section{Variational Methods}
Variational quantum algorithms encompass a broad class of methods that are among the most popular approaches to the preparation of eigenstates using current and near-term quantum hardware. 
 While the examples we consider here are optimizing a single vector, there are also many different variational methods that use subspaces \cite{2019arXiv190908925P,2022arXiv220910571F,2022arXiv221114854Y,2022PhRvA.105b2417C,2023arXiv230500060M}.  The typical strategy is a hybrid approach where the quantum device is used to prepare a parameterized family of possible wave functions, and then a classical computation is performed to minimize the associated cost function.  Let $\boldsymbol{\theta}$ be an $L$-dimensional vector of parameters $\theta_j$.  The most common example is the search for the ground state of a quantum Hamiltonian $H$ by minimizing a cost function $C({\boldsymbol \theta})$ given by the energy expectation value $C({\boldsymbol \theta}) = \braket{{\boldsymbol \theta}|H|{\boldsymbol \theta}}$ \cite{peruzzo2014variational,cerezo2021variational}.   We consider a general ansatz for the wave function $\ket{\boldsymbol{\theta}}$ that is a product of unitary operators acting upon some simple initial state $\ket{\psi_I}$ \cite{mcclean2018barren,mari2021estimating},
\begin{equation}
    \ket{{\boldsymbol \theta}} = V_L U_L(\theta_L) \cdots 
V_1 U_1(\theta_1)\ket{\psi_I}.
\end{equation}
where each $V_j$ is a fixed unitary operator.  It is convenient to take each $U_j(\theta_j)$ as an exponential of a Hermitian operator $H_j$, 
\begin{equation}
    U_j(\theta_j) = \exp(-iH_j\theta_j/2),
\end{equation}
where we restrict $H_j$ to be its own inverse so that $H_j^2 = I$.  This involutory condition is satisfied by any product of Pauli matrices on any multi-qubit system.  In such cases, we have the simple trigonometric relation,
\begin{equation}
    U_j(\theta_j) = \cos(\theta_j/2)I -i \sin(\theta_j/2)H_j.
\end{equation}
For any operator $O$, we find that 
\begin{equation}
   U_j^\dagger(\theta_j) O U_j(\theta_j) = O_1 + O_{\sin} \sin(\theta_j) + O_{\cos} \cos(\theta_j),
\end{equation}
for some operators $O_1$, $O_{\sin}$, and $O_{\cos}$ independent of $\theta_j$.
It follows that \cite{mari2021estimating}
\begin{align}
   \frac{\partial}{\partial\theta_j} U_j^\dagger(\theta_j) O U_j(\theta_j) & = O_{\sin} \cos(\theta_j) - O_{\cos} \sin(\theta_j) \nonumber \\
   & = \frac{U_j^\dagger(\theta_j+\alpha_j) O U_j(\theta_j+\alpha_j) - U_j^\dagger(\theta_j-\alpha_j) O U_j(\theta_j-\alpha_j)}{2 \sin(\alpha_j)},
\end{align}
for any $\alpha_j$ such that $\sin(\alpha_j)\ne 0$.  We note that this parameter shift formula is exact and not simply a finite-difference approximation.  This allows values for $\alpha_j$ of $O(1)$, which is helpful to measure gradient components in the presence of stochastic and systematic errors.  One can now compute the components of the gradient using
\begin{equation}
\frac{\partial}{\partial\theta_j}C(\boldsymbol{\theta} ) = \frac{C(\boldsymbol{\theta}+ \boldsymbol{\alpha}_j)-C(\boldsymbol{\theta}- \boldsymbol{\alpha}_j)}{2 \sin(\alpha_j)},
\end{equation}
where the vector $\boldsymbol{\alpha}_j$ has components $[\boldsymbol{\alpha}_j]_k = \alpha_j \delta_{jk}$.  These gradients can be used to minimize the cost function $C(\boldsymbol{\theta})$.

Consider adiabatic evolution with initial Hamiltonian $H_0$, final Hamiltonian $H_1$, and interpolating Hamiltonian $H(s) = sH_1 + (1-s)H_0$.  We then have a string of exponentials for our adiabatic evolution operator,
\begin{equation}
    U(1) = e^{-iH(1)ds} \cdots e^{-iH(s)ds} \cdots e^{-iH(0)ds},
\end{equation}
which we apply to the ground state of $H_0$.  Let $N+1$ be the number of time steps and let $ds = 1/(N+1)$.  We note that the first time step with time evolution operator $e^{-iH(0)ds}$ is usually skipped since the initial state is an eigenstate of $H_0$.  We keep here only for notational simplicity. If we now use the Trotter approximation to write 
\begin{equation}
    e^{-iH(s)ds} = e^{-i[sH_1 + (1-s)H_0]ds} \approx e^{-isH_1ds}e^{-i(1-s)H_0ds},
\end{equation}
then the adiabatic evolution operator has the form
\begin{equation}
    U(1) \approx e^{-i\gamma_NH_1ds}e^{-i\beta_NH_0ds} \cdots e^{-i\gamma_jH_1ds}e^{-i\beta_jH_0ds} \cdots e^{-i\gamma_0H_1ds}e^{-i\beta_0H_0ds},
\end{equation}
for $\beta_j = 1 - jds$ and $\gamma_j = jds$.  This structure provides the theoretical motivation for the quantum approximate optimization algorithm (QAOA) \cite{farhi2014quantum}.  Instead of using the values for $\gamma_j$ and $\beta_j$ as prescribed by adiabatic evolution, they are treated as free variational parameters optimized to minimize the energy expectation of the Hamiltonian $H_1$.

For large quantum systems, the required number of variational parameters will grow with system size.  The number of variational parameters needed as a function of the size of the system with fixed error tolerance remains an open question. 
 There are at least two major challenges that arise in quantum variational algorithms for large systems.  The first challenge is the problem of barren plateaus.  For parameterized random quantum circuits, the components of the cost function gradient will become exponentially small in the number of qubits of the quantum system \cite{mcclean2018barren}.  
 
 The second challenge is the appearance of many local minima, making gradient descent optimization difficult.  In order to quantify the difficulty of quantum variational optimization problems, we need to borrow some concepts from computational complexity theory. While the following definitions are well known in the computer science literature, they may not be familiar to researchers working on quantum many-body systems.  We therefore make a brief detour here to cover some basic definitions.
 
 A decision problem is one where the two possible answers are yes or no.  P refers to the set of decision problems that can be solved using a deterministic Turing machine in polynomial time.  NP refers to the set of decision problems whose solution, once given, can be confirmed by a deterministic Turing machine in polynomial time.  Equivalently, NP refers to decision problems that can be solved using a non-deterministic Turing machine in polynomial time, where a general non-deterministic Turing machine is endowed with the ability to branch over all possible outcomes in parallel.  A problem $p$ is NP-hard if all problems in NP can be obtained in polynomial time from the solution of $p$.  If a decision problem in NP is NP-hard, then it is called NP-complete.

Consider a graph with $d$ vertices and an adjacency matrix $A_{i,j}$ marking the edges of the graph that equal $0$ or $1$ for each pair of vertices $\{i,j\}$. The MaxCut problem poses the task of finding the subset $S$ of the vertices that maximizes the number of edges connecting $S$ and its complement, 
 \begin{equation}
 \sum_{i\in S}\sum_{j \notin S} A_{i,j}.
 \end{equation}
The MaxCut problem was shown to be NP-complete \cite{karp2010reducibility}.  The continuous MaxCut problem consists of finding the $d$-dimensional vector $\boldsymbol{\phi} = [0,2\pi)^d$ that minimizes
 \begin{equation}
 \mu({\boldsymbol{\phi}}) = \frac{1}{4}\sum_{i=1}^d \sum_{j=1}^d A_{i,j}[\cos(\phi_i)\cos(\phi_j)-1] \label{c_maxcut}.
 \end{equation}
In Ref.~{\cite{PhysRevLett.127.120502}, the continuous MaxCut problem is shown to be equivalent to the MaxCut problem and therefore also NP-hard.  Furthermore, the continuous MaxCut problem can also be recast as a variational quantum optimization problem for the Ising model Hamiltonian,
\begin{equation}
    \frac{1}{4}\sum_{i=1}^d \sum_{j=1}^d A_{i,j} (\sigma^z_i \sigma^z_j - 1), \label{Ising}
\end{equation}
with variational wave function
\begin{equation}
    e^{-i \sigma^y_d\phi_d/2} \cdots e^{-i \sigma^y_1\phi_1/2} \ket{0}^{\otimes d}.
\end{equation}
Although there is no proof that NP contains problems outside of P, there is much speculation that this is true.  NP-hard problems would then belong to the set of difficult problems outside of P, and this would include the problem of minimizing the variational cost function for an Ising Hamiltonian.  This would imply that the variational minimization problem could not be performed in polynomial time.
 
Although the general performance of variational methods for large quantum systems is challenging, there are many cases in which major simplifications arise because of some simplification, such as the emergence of a mean-field picture.  There are many examples of such approaches for fermionic quantum many-body systems \cite{kandala2017hardware, Dumitrescu:2018njn, Qian:2021wya, Kiss:2022kkz, tilly2022variational, Hlatshwayo:2022yqt, Perez-Obiol:2023vod, Sarma:2023aim}.  One popular example is the unitary coupled cluster (UCC) method.  In the UCC method, one starts with an initial state $\ket{\psi_I}$, which is a mean-field reference state.  For the unitary transformation, $U$, we take the form
\begin{equation}
 U = e^{T(\boldsymbol{\theta}) - T^\dagger(\boldsymbol{\theta})},   
\end{equation}
where 
\begin{equation}
    T(\boldsymbol{\theta}) = \sum_m T_m(\boldsymbol{\theta}),
\end{equation}
and
$T_m(\boldsymbol{\theta})$ is an $m$-body operator that produces excitations. The singles excitation has the form,
\begin{equation}
T_1(\boldsymbol{\theta}) = \sum_i \sum_a \theta^i_a a^\dagger_a a_i,
\end{equation}
where $a^\dagger_a$ and $a_i$ and fermionic creation and annihilation operators for orbitals $a$ and $i$ respectively.  The doubles terms has the structure,
\begin{equation}
T_2(\boldsymbol{\theta}) = \frac{1}{4}\sum_{i<j} \sum_{a<b} \theta^{i,j}_{a,b} a^\dagger_a a^\dagger_b a_j a_i.
\end{equation}
For the general case, we have
\begin{equation}
T_m(\boldsymbol{\theta}) = \frac{1}{(m!)^2}\sum_{i<j<\cdots} \sum_{a<b<\cdots} \theta^{i,j,\cdots}_{a,b,\cdots} a^\dagger_a a^\dagger_b \cdots a_j a_i.
\end{equation}

There are several ways to encode ferimonic antisymmetrization properties on a quantum computer.  Although often not the most efficient, the simplest approach is the Jordan-Wigner transformation \cite{jordan1928}.  We define
\begin{align}
    \sigma^+_j = & (\sigma^x_j + i \sigma^y_j)/2, \nonumber \\
    \sigma^-_j = & (\sigma^x_j - i \sigma^y_j)/2,   
\end{align}
and use the convention that $\ket{0}$ corresponds to occupation number $0$, and $\ket{1}$ corresponds to occupation number $1$.  We then have a faithful representation of the algebra of creation and annihilation operators with the mapping
\begin{align}
    a^\dagger_j & = \sigma^-_j \otimes \sigma^z_{j-1} \otimes \cdots \otimes \sigma^z_{1}, \nonumber \\
    a_j & =  \sigma^+_j \otimes \sigma^z_{j-1} \otimes \cdots \otimes \sigma^z_{1}.  
\end{align}
This gives the required anticommutation relations,
\begin{equation}
    \{a_j,a^\dagger_k\} = \delta_{j,k}, \; \;
    \{ a_j,a_k \} = \{ a^\dagger_j,a^\dagger_k \} = 0.
\end{equation}
Many other antisymmetrization techniques \cite{Bravyi:2002, verstraete2005mapping, whitfield2016local, steudtner2018fermion, derby2021compact, nys2023quantum} have been designed that are computationally more efficient in cases where the products of creation and annihilation operators in the Hamiltonian appear in combinations with some locality restriction with respect to the orbital index.

A convenient choice for the mean-field reference state $\ket{\psi_I}$ is a Hartree-Fock state, corresponding to a Slater determinant of single-particle orbitals achieving the lowest energy expectation value.  The Thouless theorem \cite{thouless1960stability} shows how to prepare any desired Slater determinant state starting from any other Slater determinant state.  Let $\alpha_p({\bf r})$ label the original orbitals and let $\beta_p({\bf r})$ label the new orbitals.  We take $a^\dagger_p, a_p$ to be the creation and annihilation operators for $\alpha_p({\bf r})$, and $b^\dagger_p, b_p$ to be the creation and annihilation operators for $\beta_p({\bf r})$.  
Let $N$ be the number of particles in our system of interest. The aim is to derive a simple relation between $b^\dagger_N \cdots b^\dagger_1 \ket{\rm vac}$ and $a^\dagger_N \cdots a^\dagger_1 \ket{\rm vac}$.  Without loss of generality, we use a linear transformation to redefine the orbitals $\beta_1({\bf r}), \cdots, \beta_N({\bf r})$ so that for each $p = 1, \cdots, N$, we have
\begin{equation}
    b^\dagger_p = a^\dagger_p + \sum_{q=N+1}^\infty a^\dagger_q u_{q,p}
\end{equation}
for some coefficient matrix $u_{q,p}$.  
The linear transformation on $\beta_1({\bf r}), \cdots, \beta_N({\bf r})$ has no effect on $b^\dagger_N \cdots b^\dagger_1 \ket{\rm vac}$ except for introducing an overall normalization factor.  Our convention will ensure that $b^\dagger_N \cdots b^\dagger_1 \ket{\rm vac}$ and $a^\dagger_N \cdots a^\dagger_1 \ket{\rm vac}$ have the same normalization.

The Thouless theorem is based on the observation that for each $p = 1, \cdots, N$, 
\begin{align}
 & \left( a^\dagger_p +  \sum_{q=N+1}^\infty a^\dagger_q u_{q,p} \right) F({\rm no}\; a^\dagger_p) \ket{\rm vac} \nonumber \\
& = \left( 1 + \sum_{q=N+1}^\infty a^\dagger_q  u_{q,p} a_p \right) a^\dagger_p F({\rm no}\; a^\dagger_p) \ket{\rm vac},
\end{align}
where $F({\rm no}\; a^\dagger_p)$ is an arbitrary function of the creation and annihilation operators where $a^\dagger_p$ does not appear.  We then have 
\begin{align}
    b^\dagger_N \cdots b^\dagger_1 \ket{\rm vac} & = \left( a^\dagger_N +  \sum_{q=N+1}^\infty a^\dagger_q u_{q,N} \right) \cdots \left( a^\dagger_1 +  \sum_{q=N+1}^\infty a^\dagger_q u_{q,1} \right) \ket{\rm vac} \nonumber \\
    & = \left( 1 +  \sum_{q=N+1}^\infty a^\dagger_q u_{q,N} a_N \right) a^\dagger_N \cdots \left( 1 +  \sum_{q=N+1}^\infty a^\dagger_q u_{q,1} a_1 \right) a^\dagger_1 \ket{\rm vac}. 
\end{align}
This leads to the simple relation,
\begin{align}
    b^\dagger_N \cdots b^\dagger_1 \ket{\rm vac} & = \left( 1 +  \sum_{q=N+1}^\infty a^\dagger_q u_{q,N} a_N \right) \cdots \left( 1 +  \sum_{q=N+1}^\infty a^\dagger_q u_{q,1} a_1 \right) a^\dagger_N \cdots a^\dagger_1 \ket{\rm vac} \nonumber \\
    & = \exp \left( \sum_{p=1}^N \sum_{q=N+1}^\infty a^\dagger_q u_{q,p} a_p \right) a^\dagger_N \cdots a^\dagger_1 \ket{\rm vac}. \label{Thouless}
\end{align}
Once the Hartree-Fock orbitals are determined using classical computing, one can prepare a simple $N$-particle Slater determinant state with orbitals given by the computational basis of the quantum computer and then apply the transformation in Eq.~\eqref{Thouless} \cite{doi:10.1126/science.abb9811}.

\section{Phase Detection Algorithms}
Quantum phase estimation \cite{Kitaev:1995qy} is a well-known example of a phase detection algorithm that can be used to find energy eigenvalues and prepare energy eigenstates of the quantum many-body problem \cite{Abrams:1997gk,dorner2009optimal,svore2013faster,RuizGuzman:2021qyj}. Suppose for the moment that $\ket{\psi}$ is an eigenstate of the unitary operator $U$ with eigenvalue $e^{2\pi i\theta}$.  Of particular interest is the case where the unitary operator $U$ is the time evolution operator for some Hamiltonian $H$ over some fixed time step $\Delta t$.  The goal is to efficiently determine the phase angle $\theta$.  Since $U\ket{\psi}=e^{2\pi i \theta}\ket{\psi}$, we have $U^{2^j}\ket{\psi} = e^{2\pi i \theta 2^j}\ket{\psi}$ for any nonnegative integer $j$.  Together with the state $\ket{\psi}$, we take $n$ ancilla qubits with each initialized as $\ket{0}$.  The resulting state is $\ket{0}^{\otimes n} \otimes \ket{\psi}$.  The Hadamard gate is a single qubit gate that maps $\ket{0}$ to $1/\sqrt{2}(\ket{0} + \ket{1})$ and maps $\ket{1}$ to $1/\sqrt{2}(\ket{0} - \ket{1})$.  The action of the Hadamard gate for a general linear combination of $\ket{0}$ and $\ket{1}$ is determined by linearity.  We apply Hadamard gates to each of the ancilla qubits so that we get
\begin{equation}
    \frac{1}{2^{n/2}}\left( \ket{0} + \ket{1} \right)^{\otimes n} \otimes \ket{\psi}.
\end{equation}
For each of the ancilla qubits $j = 0, \cdots, n-1$, we use the ancilla qubit to control the unitary gate $U^{2^j}$.  This means that $U^{2^j}$ is applied when the ancilla qubit $j$ is in state $\ket{1}$, but no operation is performed if the ancilla qubit is in state $\ket{0}$.  The result we get is \cite{nielsen2010quantum} 
\begin{equation}
\frac{1}{2^{n/2}}\left( \ket{0} + e^{2\pi i \theta 2^{n-1}}\ket{1}\right) \otimes \cdots \otimes \left( \ket{0} + e^{2\pi i \theta 2^{0}}\ket{1}\right) \otimes \ket{\psi} = \ket{f(\theta)} \otimes \ket{\psi},
\end{equation}
where
\begin{align}
\ket{f(\theta)} = & \frac{1}{2^{n/2}}\sum_{m=0}^{2^n-1} \left( e^{2\pi i \theta 2^{n-1} m_{n-1}} \ket{m_{n-1}} \right) \otimes \cdots \left(e^{2\pi i \theta 2^0 m_{0}} \otimes \ket{m_0} \right) \nonumber \\
= & \frac{1}{2^{n/2}}\sum_{m=0}^{2^n-1} e^{2\pi i \theta m} \ket{m_{n-1}} \otimes \cdots \otimes \ket{m_0},
\end{align}
and $m_{n-1}\cdots m_0$ are the binary digits of the integer $m$.  

Let $k$ be an integer between $0$ and $2^n-1$ with binary representation $k_{n-1} \cdots k_0$.  We note that when $\theta$ equals $k$ divided by $2^n$, then $\ket{f(k/2^n)}$ is the quantum Fourier transform of the state $\ket{k_{n-1}} \otimes \cdots \otimes \ket{k_0}$, 
\begin{equation}
\ket{f(k/2^n)} = \frac{1}{2^{n/2}}\sum_{m=0}^{2^n-1} e^{2\pi i k m/2^n} \ket{m_{n-1}} \otimes \cdots \otimes \ket{m_0}.
\end{equation}
We can therefore extract information about the value of $\theta$ by applying the inverse quantum Fourier transform to $\ket{f(\theta)}$,
\begin{equation}
    QFT^{-1}\ket{f(\theta)} = \frac{1}{2^{n}}\sum_{k=0}^{2^n-1}\sum_{m=0}^{2^n-1} e^{-2\pi i (k/2^n-\theta) m} \ket{k_{n-1}} \otimes \cdots \otimes \ket{k_0}.
\end{equation}
We see that if $\theta$ equals $k/2^n$ for some integer $k$ in the summation, then $QFT^{-1}\ket{f(\theta)}$ equals $\ket{k_{n-1}} \otimes \cdots \otimes \ket{k_0}$.  In the general case, we get a superposition of such states $\ket{k_{n-1}} \otimes \cdots \otimes \ket{k_0}$ that is highly peaked for integers $k$, where $k/2^n$ is close to $\theta$.  We simply measure each ancilla qubit and determine $k/2^n$ to obtain an estimate of $\theta$.  This is repeated over several trials to build a probability distribution and refine the estimate of $\theta$. 

Suppose now that $\ket{\psi}$ is not an eigenstate of $U$ but rather a general superposition of eigenstates $\ket{\psi_a}$ with eigenvalues $e^{2\pi i\theta_a}$,
\begin{equation}
    \ket{\psi} = \sum_a c_a \ket{\psi_a}.
\end{equation}
We can now apply phase estimation to the general state $\ket{\psi}$ in exactly the same manner as before.  Let us assume that the separation between each $\theta_a$ is large compared to $1/2^n$.  This ensures that the peaked distributions we get for each eigenvector have negligible overlap.  The outcome after measuring the $n$ ancilla qubits will be 
\begin{equation}
    \ket{k_{n-1}} \otimes \cdots \otimes \ket{k_0} \otimes \ket{\psi_a},
\end{equation}
for some eigenstate $\ket{\psi_a}$.  The probability of $\ket{\psi_a}$ being selected will equal $|c_a|^2$.  The error of quantum phase estimation in determining eigenvalues will scale inversely with $2^n$.  This arises from the discretization of energy values $k/2^n$, where $k$ is an integer from $0$ to $2^n-1$.  If we relate $U$ to the time evolution of a Hamiltonian $H$ for time step $\Delta t$, the error in the energy scales inversely with the total time evolution required.  This scaling of the uncertainty matches the lower bound one expects from the Heisenberg uncertainty principle.  
The error of phase estimation for eigenstate preparation arises from the admixture of terms from different eigenstates,
\begin{equation}
   \frac{1}{2^n} \sum_a \sum_{m=0}^{2^n-1} c_a e^{-2\pi i (k/2^n-\theta_a) m} \ket{k_{n-1}} \otimes \cdots \otimes \ket{k_0} \otimes \ket{\psi_a}.
\end{equation}
When the spacing between $\theta_a$ is much larger than $1/2^n$, then the contamination of other eigenstates will be $O(2^{-n})$.  For the case when $U$ is the time evolution of a Hamiltonian $H$ for time duration $\Delta t$, then the error of eigenstate preparation scales inversely with the total amount of time evolution needed.

We have mentioned the quantum Fourier transform, but have not yet discussed how it is implemented.  It suffices to describe its action on the state $\ket{k_{n-1}} \otimes \cdots \otimes \ket{k_0}$.  We again use the notation that $k_{n-1} \cdots k_0$ are the binary digits of the integer $k$.  The desired action of the quantum Fourier transform upon $\ket{k_{n-1}} \otimes \cdots \otimes \ket{k_0}$ is
\begin{align}
    \frac{1}{2^{n/2}} \left( \ket{0} + e^{2\pi i k 2^{n-1}/ 2^n} \ket{1} \right) & \otimes \cdots \otimes \left( \ket{0} + e^{2\pi i k 2^0/ 2^n} \ket{1} \right) \nonumber \\
    & = \frac{1}{2^{n/2}}\sum_{m=0}^{2^n-1} e^{2\pi i k m/2^n} \ket{m_{n-1}} \otimes \cdots \otimes \ket{m_0}. 
\end{align}
The first few steps of the quantum Fourier transform algorithm will actually produce the desired result with the tensor product in the reverse order,
\begin{equation}
    \frac{1}{2^{n/2}} \left( \ket{0} + e^{2\pi i k 2^0/ 2^n} \ket{1} \right) \otimes \cdots \otimes  \left( \ket{0} + e^{2\pi i k 2^{n-1}/ 2^n} \ket{1} \right) \label{QFTreverse}.
\end{equation}
But this can be fixed by pairwise swap gates between qubits $0$ and $n-1$, $1$ and $n-2$, etc.

The quantum Fourier transform begins with the state
\begin{equation}
    \ket{k_{n-1}} \otimes \ket{k_{n-2}} \otimes \cdots \otimes \ket{k_0}.
\end{equation}
We first act upon qubit $n-1$ with a Hadamard gate and this gives
\begin{equation}
    \frac{1}{2^{1/2}}\left( \ket{0} + e^{2\pi i k_{n-1} 2^{n-1}/ 2^n}\ket{1} \right) \otimes \ket{k_{n-2}} \otimes \cdots \otimes \ket{k_0}.
\end{equation}
The coefficient in front of $\ket{1}$ equals $1$ if $k_{n-1}=0$ and equals $-1$ if $k_{n-1}=1$.  We use qubit $n-2$ to apply a controlled phase rotation to qubit $n-1$ by a phase $e^{2\pi i k_{n-2}2^{n-2}/2^n}$.  The result is
\begin{equation}
    \frac{1}{2^{1/2}}\left( \ket{0} + e^{2\pi i (k_{n-1} 2^{n-1} + k_{n-2} 2^{n-2})/ 2^n}\ket{1} \right) \otimes \ket{k_{n-2}} \otimes \cdots \otimes \ket{k_0}.
\end{equation}
We continue in this manner with qubit $j$ applying a controlled phase rotation on qubit $n-1$ by a phase $e^{2\pi i k_{j}2^{j}/2^n}$.  After doing this for all of the remaining qubits, we get
\begin{equation}
    \frac{1}{2^{1/2}}\left( \ket{0} + e^{2\pi i k/ 2^n}\ket{1} \right) \otimes \ket{k_{n-2}} \otimes \cdots \otimes \ket{k_0}.
\end{equation}
We perform the analogous process for qubits $n-2, \cdots, 1$.  For the qubit $0$, we simply apply the Hadamard gate. In the end, we get the desired result, 
\begin{equation}
    \frac{1}{2^{n/2}} \left( \ket{0} + e^{2\pi i k 2^0/ 2^n} \ket{1} \right) \otimes \cdots \otimes  \left( \ket{0} + e^{2\pi i k 2^{n-1}/ 2^n} \ket{1} \right).
\end{equation}
As described above, we now apply swap gates between pairs of qubits $0$ and $n-1$, $1$ and $n-2$, etc. and then we obtain the desired quantum Fourier transform.

Iterative phase estimation performs the determination of the binary digits of $\theta$ one at a time \cite{Kitaev:1995qy,dobvsivcek2007arbitrary,o2009iterative,lanyon2010towards,wiebe2016efficient,ding2023even}.  Let $\ket{\psi}$ again be an eigenstate of $U$ with eigenvalue $e^{2\pi i \theta}$. 
 We first consider the case where $\theta$ is equal to $k/2^n$ where $k$ is an integer between $0$ and $2^n-1$.  We start with $\ket{0}\otimes \ket{\psi}$ and apply a Hadamard gate to obtain
 \begin{equation}
     \frac{1}{2^{1/2}}\left(\ket{0} + \ket{1} \right)\otimes \ket{\psi}.
 \end{equation}
We now use the ancilla qubit to perform the controlled unitary operator $U^{2^{n-1}}$.  The result is then
\begin{equation}
    \ket{f_0(\theta)} \otimes \ket{\psi},
\end{equation}
where
\begin{equation}
      \ket{f_0(\theta)} = \frac{1}{2^{1/2}}\left(\ket{0} + e^{2\pi i \theta 2^{n-1}} \ket{1} \right).
\end{equation}
Applying a Hadamard gate to $\ket{f_0(\theta)}$ gives 
\begin{align}
     \left( \frac{1 + e^{2\pi i \theta 2^{n-1}}}{2}
      \ket{0} + \frac{1 - e^{2\pi i \theta 2^{n-1}}}{2}
      \ket{1} \right) = \delta_{k_0,0}\ket{0} + \delta_{k_0,1}\ket{1}.
\end{align}
Therefore, we can determine the digit $k_0$.  We note that in iterative phase estimation, the digit $k_0$ is determined using controlled time evolution for $U^{2^{n-1}}$ rather than $U^{2^0}$. This is because $k_0$ is the remainder we get when $\theta 2^n$ is divided by $2$.  Therefore $e^{2\pi i \theta 2^{n-1}}$ is either $1$ for $k_0=0$ or $-1$ for $k_0=1$.

Let us assume that we have determined the digits from $k_0, \cdots, k_{j-1}$.  We can determine $k_j$ by taking 
\begin{equation}
     \frac{1}{2^{1/2}}\left(\ket{0} + \ket{1} \right)\otimes \ket{\psi}
 \end{equation}
and using the ancilla qubit to perform the controlled unitary operator $U^{2^{n-j-1}}$ followed by the phase gate \\[2pt]
\begin{equation}
   \ket{0}\bra{0} + e^{- 2 \pi i (k_{j-1}2^{-2} + \; \cdots \; + k_{0}2^{-j-1})}\ket{1}\bra{1}, \\[8pt]
\end{equation}
on the ancilla qubit.  This phase gate removes the complex phase associated with the binary digits $k_{0}, \cdots, k_{j-1}$ that have already been determined. The net result is 
\begin{equation}
    \ket{f_j(\theta)} \otimes \ket{\psi},
\end{equation}
where \\[2pt]
\begin{equation}
      \ket{f_j(\theta)} = \frac{1}{2^{1/2}}\left(\ket{0} + e^{2\pi i \theta 2^{n-j-1} - 2 \pi i (k_{j-1}2^{-2} + \; \cdots \; + k_{0}2^{-j-1}) } \ket{1} \right). \\[8pt]
\end{equation}
Applying a Hadamard gate to $\ket{f_j(\theta)}$ gives 
\begin{align}
\delta_{k_j,0}\ket{0} + \delta_{k_j,1}\ket{1}.
\end{align}

For the general case where $\theta$ is not equal to $k/2^n$ for some integer $k$ between $0$ and $2^n-1$, there will be some distribution of values associated with the measurements of the binary digits $k_{n-1}, \cdots, k_0$.  As with regular phase estimation, the error in energy resolution scales inversely with $2^n$ and is therefore inversely proportional to the number of operations of $U$ needed.  If $U$ is the time evolution of a Hamiltonian $H$ over time step $\Delta t$, then the error in the energy scales inversely with the total time evolution required. Iterative phase estimation is not designed to perform eigenstate preparation.  If we start from a general linear combination of energy eigenstates, then the uncertainty in the sequential measurements of the binary digits $k_{n-1}, \cdots, k_0$ arising from the different eigenvalues $e^{2\pi i \theta_a}$ will prevent the algorithm from functioning as intended.

The rodeo algorithm is another phase detection algorithm \cite{Choi:2020pdg,Bee-Lindgren:2022,Qian:2021jxp} that shares some structural similarities with iterative phase estimation.  In contrast to iterative phase estimation, however, the rodeo algorithm is efficient in preparing energy eigenstates starting from a general initial state.  Let $H$ be the Hamiltonian for which we want to prepare energy eigenstates.  To explain the algorithm, we first consider the case where the initial state is an eigenstate of $H$.  We call it $\ket{\psi_j}$ with eigenvalue $E_j$.  We use one ancilla qubit and start with the state 
\begin{equation}
    \ket{0} \otimes \ket{\psi_j},
\end{equation}
and apply the Hadamard gate on the ancilla qubit,
\begin{equation}
    \frac{1}{2^{1/2}}\left( \ket{0} + \ket{1} \right) \otimes \ket{\psi_j}.
\end{equation}
We then use the ancilla to perform the controlled unitary for $e^{-i H t_1}$ and apply the phase gate \\[2pt]
\begin{equation}
   \ket{0}\bra{0} + e^{iEt_1}\ket{1}\bra{1}, \\[8pt]
\end{equation}
on the ancilla.  This produces
\begin{equation}
    \frac{1}{2^{1/2}}\left( \ket{0} + e^{-i(E_j-E)t_1}\ket{1} \right) \otimes \ket{\psi_j}.
\end{equation}
We now apply a Hadamard gate to the ancilla qubit, which then gives
\begin{equation}
    \frac{1}{2}\left[ \left( 1 + e^{-i(E_j-E)t_1} \right) \ket{0}+ \left( 1 - e^{-i(E_j-E)t_1} \right) \ket{1} \right] \otimes \ket{\psi_j}.
\end{equation}
If we measure the ancilla qubit, the probability of measuring $\ket{0}$ is $\cos^2[(E_j-E)t_1/2]$ and the probability of measuring $\ket{1}$ is $\sin^2[(E_j-E)t_1/2]$.  We call the measurement of $\ket{0}$ a success and the measurement of $\ket{1}$ a failure.  We repeat this process for $n$ cycles with times $t_1, \cdots, t_n$.  The probability of success for all $n$ cycles is 
\begin{equation}
    \prod_{k=1}^n \cos^2[(E_j-E)t_k/2].
\end{equation}
If we take random times $t_1, \cdots, t_n$ to be chosen from a Gaussian normal distribution with zero mean and $\sigma$ standard deviation, then the success probability averaged over many trials will equal
\begin{equation}
    P_n(E) =  \frac{\left[1+e^{-(E_j-E)^2\sigma^2/2}\right]^n  }{2^n}.
\end{equation}
We see that the peak value is equal to $1$ when $E_j = E$ and the width of the peak is $O(\sigma^{-1}n^{-1/2}).$

Let us now consider a general linear combination of energy eigenstates
\begin{equation}
    \ket{\psi} = \sum_j c_j \ket{\psi_j}. 
\end{equation}
For this case, the probability of success for $n$ cycles is
\begin{equation}
    P_n(E) = \sum_j  \frac{\left[1+e^{-(E_j-E)^2\sigma^2/2}\right]^n \left| c_j \right|^2 }{2^n}.
\end{equation}
When scanning over the input parameter $E$, peaks in $P_n(E)$ will appear at places where there are eigenvalues $E_j$ and the overlap with the initial state is not too small.  For fixed $n$, the error of the energy determination scales inversely with $\sigma$.  Similarly to phase estimation and iterative phase estimation, the rodeo algorithm saturates the Heisenberg bound, where the error in the energy scales inversely with the total duration of time evolution.  

In contrast with both phase estimation and iterative phase estimation, the rodeo algorithm is exponentially fast for eigenstate preparation.  There are several other energy projection and filtering methods with similar characteristics \cite{ge2019faster,lu2021algorithms,Stetcu:2022nhy}. Once the peak of the eigenstate energy in $P_n(E)$ is located approximately, we set $E$ as the peak value.  With $E$ fixed and $\sigma$ fixed, the error estimates for the eigenvector scale as $1/2^n$ for small $n$ and accelerate to $1/4^n$ for asymptotically large values of $n$  \cite{Choi:2020pdg}.  The $1/2^n$ comes from the fact that the arithmetic mean of $\cos^2(\theta)$ equals $1/2$, while the $1/4^n$ comes from the fact that the geometric mean of $\cos^2(\theta)$ equals $1/4$.  In Ref.~\cite{Cohen:2023rhd}, it was shown that the use of progressive smaller values for the time evolution parameters $t_j$ accelerates the convergence of the rodeo algorithm towards $1/4^n$.  The main limitation of the rodeo algorithm for large quantum many-body systems is the requirement that the initial state have non-negligible overlap with the eigenstate of interest.  This is a difficult problem that is common to nearly all eigenstate preparation algorithms that use measurement projection. Nevertheless, one can use techniques such as adiabatic evolution, variational methods, or some other approach as a preconditioner to significantly increase the overlap with the eigenstate of interest \cite{Choi:2020pdg}. 

\section{Summary and Outlook}
In this article, we have presented several methods that show the essential features of adiabatic evolution, variational methods, and phase detection algorithms.  All of the algorithms have their strengths and limitations, and one common theme is that the techniques can be combined with each other to produce something that is potentially greater than the sum of its parts.  For example, adiabatic evolution provides a theoretical foundation for the QAOA variational method.  In turn, the variational method can be used to find a good starting Hamiltonian for adiabatic evolution.  Both adiabatic evolution and variation methods can be used as an initial-state preconditioner for phase detection algorithms.

There has been great interest by both scientists and the general public on the question of quantum advantage, if and when quantum computers are able to perform tasks exceeding the capabilities of classical computers.  It is generally believed that calculations of real-time dynamics and spectral functions of quantum many-body systems are areas ripe for possible quantum advantage.  However, the dynamics of some quantum many-body system starting from a trivial initial state is not something that connects directly with real-world phenomena.  To make connections with real-world experiments and observations, one also needs the ability to prepare energy eigenstates.  It is not clear whether quantum advantage will be achievable for the task of eigenstate preparation.  However, this may not be necessary.  It may be enough for quantum eigenstate preparation to be competitive with classical computing methods to achieve quantum advantage for calculating the real-time dynamics or spectral functions for real-world applications. The algorithms described in this article provide some of the tools needed, but much more work is needed to address the remaining challenges. 

\bmhead{Acknowledgments}
The author acknowledges support from the U.S. National Science Foundation (PHY-2310620), the U.S. Department of Energy (DE-SC0021152, DE-SC0013365, DE-SC0023658) and the SciDAC-4 and SciDAC-5 NUCLEI Collaborations.  This research used resources of the Oak Ridge Leadership Computing Facility, which is a DOE Office of Science User Facility supported under Contract DE-AC05-00OR22725.






\bibliography{References}

\end{document}